\documentclass[10pt, twocolumn]{IEEEtran}
\usepackage{amsmath}
\usepackage{amssymb}
\usepackage{mathrsfs}
\usepackage{cite}
\usepackage{epsfig}
\usepackage{epsf}
\usepackage{theorem}
\usepackage{graphics}
\usepackage[active]{srcltx}

\newtheorem{defi}{Definition}

\begin{document}

\title{On the Energy Efficiency of LT Codes in\\
Proactive Wireless Sensor Networks
\thanks{Manuscript received April 1, 2010; revised June 28, 2010 and September 22, 2010;
accepted November 9, 2010. The associate editor coordinating the review of this
manuscript and approving it for publication was Prof. Shengli Zhou.}
\thanks{Copyright (c) 2010 IEEE. Personal use of this material is permitted.
However, permission to use this material for any other purposes must be obtained
from the IEEE by sending a request to pubs-permissions@ieee.org.}
\thanks{J. Abouei is with the Department of Electrical Engineering,
Yazd University, Yazd, Iran (e-mail: abouei@yazduni.ac.ir). }
\thanks{J. D. Brown is with the DRDC Ottawa, Ottawa, ON, Canada. He completed his
Ph.D. in ECE at University of Toronto (e-mail:
david\_jw\_brown@yahoo.com).}
\thanks{K. N. Plataniotis, and S. Pasupathy are with the Edward S.
Rogers Sr. Department of Electrical and Computer Engineering,
University of Toronto, Toronto, ON M5S 3G4, Canada (e-mails:
\{kostas, pas\}@comm.utoronto.ca).}
\thanks{The material in this paper was presented
in part at the QBSC'2010 conference, Kingston, Canada, May 2010 \cite{Jamshid_BSC2010}.
The work of J. Abouei was performed when he was with the Department of Electrical
and Computing Engineering, University of Toronto, Toronto, ON M5S 3G4, Canada.
The work was supported in part by an Ontario Research Fund (ORF) project
entitled ``Self-Powered Sensor Networks''.}}

\author{\small Jamshid Abouei, \emph{Member, IEEE,} J. David
Brown, Konstantinos N. Plataniotis, \emph{Senior Member, IEEE} and \\
Subbarayan Pasupathy, \emph{Life Fellow, IEEE}}

\maketitle

\markboth{IEEE Transactions on Signal Processing}{}

\begin{abstract}
This paper presents an in-depth analysis on the energy efficiency of
Luby Transform (LT) codes with Frequency Shift Keying (FSK)
modulation in a Wireless Sensor Network (WSN) over Rayleigh fading
channels with path-loss. We describe a proactive system model
according to a flexible duty-cycling mechanism utilized in practical
sensor apparatus. The present analysis is based on realistic
parameters including the effect of channel bandwidth used in the
IEEE 802.15.4 standard, active mode duration and computation energy.
A comprehensive analysis, supported by some simulation studies on
the probability mass function of the LT code rate and coding gain,
shows that among uncoded FSK and various classical channel coding
schemes, the optimized LT coded FSK is the most energy-efficient
scheme for distance $d$ greater than the pre-determined threshold
level $d_T$, where the optimization is performed over coding and
modulation parameters. In addition, although the optimized uncoded
FSK outperforms coded schemes for $d < d_T$, the energy gap between
LT coded and uncoded FSK is negligible for $d < d_T$ compared to the
other coded schemes. These results come from the flexibility of the
LT code to adjust its rate to suit instantaneous channel conditions,
and suggest that LT codes are beneficial in practical low-power WSNs
with dynamic position sensor nodes.
\end{abstract}


\section{Introduction}
Wireless Sensor Networks (WSNs) have been recognized as a new
generation of ubiquitous computing systems to support a broad range
of applications, including monitoring, health care and tracking
environmental pollution levels. Minimizing the total energy
consumption in both circuit components and RF signal transmission is
a crucial challenge in designing a WSN. Central to this study is to
find energy-efficient modulation and coding schemes in the physical
layer of a WSN to prolong the sensor lifetime
\cite{Cui_GoldsmithITWC0905, HowardEURASIP2006}. For this purpose,
energy-efficient modulation/coding schemes should be simple enough
to be implemented by state-of-the-art low-power technology, but
still robust enough to provide the desired service. Furthermore,
since sensor devices frequently switch from sleep mode to active
mode, modulation and coding circuits should have fast start-up times
\cite{Wang_ISLPED2001} along with the capability of transmitting
packets during a pre-assigned time slot before new sensed packets
arrive. In addition, a WSN needs a powerful channel coding scheme
(when the distance between nodes exceeds a certain threshold level)
to protect transmitted data against the unpredictable and harsh
nature of channels. We refer to these low-complexity and low-energy
consumption approaches in WSNs providing proper link reliability as
\emph{Green Modulation/Coding} (GMC) schemes.

There have been several recent works on the energy efficiency of
various modulation/ coding schemes in WSNs (see e.g.,
\cite{Cui_GoldsmithITWC0905, TangITWC0407, Chouhan_ITWC1009}). Tang
\emph{et al.} \cite{TangITWC0407} compare the power efficiency of
PPM and FSK in a WSN over fading channels with path-loss without
considering the effect of channel coding. Reference
\cite{Sankarasubramaniam2003} investigates the energy efficiency of
BCH and convolutional codes with FSK for the optimal packet length
in a point-to-point WSN. It is shown in
\cite{Sankarasubramaniam2003} that BCH codes can improve energy
efficiency compared to the convolutional code for the optimal fixed
packet size. Liang \emph{et al.} \cite{LiangPACRIM2007} investigate
the energy efficiency of uncoded FSK modulation in a WSN, where
multiple senders transmit their data to a central node in a
Frequency-Division Multiple Access (FDMA) fashion. Reference
\cite{Hanzo_2009} presents the hardware implementation of the
Forward Error Correction (FEC) encoder in IEEE 802.15.4 WSNs, which
employs parallel processing to achieve a low processing latency and
energy consumption.

Most of the pioneering work on energy-efficient modulation/coding,
including research in \cite{TangITWC0407,
Sankarasubramaniam2003}, has focused only on
minimizing the energy consumption of transmitting one bit, ignoring
the effect of bandwidth and transmission time duration. In a
practical WSN however, it is shown that minimizing the total energy
consumption depends strongly on the active mode duration and the
channel bandwidth. References \cite{Cui_GoldsmithITWC0905},
\cite{Chouhan_ITWC1009} and \cite{JamshidICASSP_2010} address this
issue in a point-to-point WSN, where a sensor node transmits an
equal amount of data per time unit to a designated sink node. In
\cite{Cui_GoldsmithITWC0905}, the authors show that uncoded MQAM is
more energy-efficient than uncoded MFSK for short-range
applications, while using convolutional coded MFSK over AWGN is
desirable for longer distances. This line of work is further
extended in \cite{Chouhan_ITWC1009} by evaluating the energy
consumption per information bit of a WSN for Reed Solomon (RS) Codes
and various modulation schemes over AWGN channels with path-loss. In
\cite{Cui_GoldsmithITWC0905} and \cite{Chouhan_ITWC1009}, the
authors do not consider the effect of multi-path fading.

More recently, the attention of researchers has been drawn to
deploying rateless codes (e.g., Luby Transform (LT) code
\cite{LubyFOCS2002}) in WSNs due to the significant advantages of
these codes in erasure channels. For instance in
\cite{Eckford_ICC2006}, the authors present a scheme for cooperative
error control coding using rateless and Low-Density Generator-Matrix
(LDGM) codes in a multiple relay WSN. However, investigating the
energy efficiency of rateless codes in WSNs with low-energy
modulations over realistic fading channel models has received little
attention. To the best of our knowledge, there is no existing
analysis on the energy efficiency of rateless coded modulation that
considers the effect of channel bandwidth and active mode duration
on the total energy consumption in a proactive WSN. This paper
addresses this issue and presents the first in-depth analysis of the
energy efficiency of LT codes with FSK, known as green modulation as
described in \cite{JamshidICASSP_2010}. The present analysis is
based on a realistic model in proactive WSNs operating over a
Rayleigh fading channel with path-loss. In addition, we obtain the
probability mass function of the LT code rate and the corresponding
coding gain, and study their effects on the energy efficiency of the
WSN. This study uses the classical BCH and convolutional codes (as
reference codes), utilized in IEEE standards, for comparative
evaluation. Numerical results, supported by some experimental setup
on the computation energy, show that the optimized LT coded FSK
scheme is the most energy-efficient scheme for distance $d$ greater
than the threshold level $d_T$. In addition, although the optimized
uncoded FSK outperforms coded schemes for $d < d_T$, the energy gap
between LT coded and uncoded FSK is negligible for $d < d_T$
compared to the other coded schemes. This result comes from the
simplicity and flexibility of the LT codes, and suggests that LT
codes are beneficial in practical low-power WSNs with dynamic
position sensor nodes.

The rest of the paper is organized as follows. In Section
\ref{System_model}, the proactive system model over a realistic
wireless channel is described. The energy consumption of both
circuits and signal transmission of uncoded MFSK modulation is
analyzed in Section \ref{uncoded_MFSK}. Design of LT codes and the
energy efficiency of the LT coded MFSK are presented in Section
\ref{analysis_Ch4}. Section \ref{simulation_Ch5} provides some
numerical evaluations using some classical channel codes as well as
realistic models to confirm our analysis. Also, some design
guidelines for using LT codes in practical WSN applications are
presented. Finally in Section \ref{conclusion_Ch6}, an overview of
the results and conclusions are presented.

For convenience, we provide a list of key mathematical symbols used
in this paper in Table I. For simplicity of notation, we use the
superscripts `BC', `CC' and `LT' for BCH, convolutional and LT
codes, respectively. We use $\mathcal{E}$, $\mathcal{P}$ and $T$ for
energy, power and time parameters, respectively. Also, we use the
subscript ``$c$'' to distinguish coding parameters from the uncoded
ones.

\begin{table}
   \label{table234}
\caption{List of Notations } \centering
  \begin{tabular}{|l|l|}
  \hline
  $B$:~Channel bandwidth                   & $b$:~Number of bits in each symbol\\
  $d$:~Transmission distance               & $\mathcal{E}_t$:~Energy of transmitted signal\\
  $\mathbb{E}[~.~]$:~Expectation operator  & $\mathcal{E}_{N}$:~Total energy consumption\\
  $M$:~Constellation size                  & $h_{i}$:~Fading channel coefficient\\
  $N$:~Number of sensed message            & $\mathcal{L}_d$:~Path loss gain\\
  $n$:~Codeword block length               & $\mathcal{O}(x)$:~Output-node degree distribution\\
  $P_b$:~Bit error rate                    & $\mathcal{P}_c$:~Circuit power consumption\\
  $P_R(\ell)$:~pmf of LT code rate         & $\mathcal{P}_t$:~Power of transmitted signal\\
  $R_c$:~Code rate                         & $T_{ac}$:~Active mode duration\\
  $\eta$:~Path-loss exponent               & $T_{tr}$:~Transient mode duration\\
  $\Omega=\mathbb{E}\left[\vert h_{i} \vert^2\right]$   & $T_s$:~Symbol duration\\
  $\Upsilon_{c}$:~Coding gain              & $\gamma_{i}$:~Instantaneous SNR \\
  \hline
  \end{tabular}
\end{table}

\section{System Model and Assumptions}\label{System_model}
We consider a proactive wireless sensor system, in which a sensor
node transmits an equal amount of data per time unit to a designated
sink node. Such a proactive sensor system is typical of many
environmental applications such as sensing temperature, humidity and
level of contamination \cite{Cordeiro_Book2006}. We assume
a non real-time service application where the data transmission between the
sensor and the sink nodes does not have tight
constraint on delay. The sensor and sink
nodes synchronize with one another and operate in a real-time based
process as depicted in Fig \ref{fig: Time-Basis}. During
\emph{active mode} period $T_{ac}$, the sensed analog signal is
first digitized by an Analog-to-Digital Converter (ADC), and an
$N$-bit message sequence $\mathcal{M}_N\triangleq (m_1,m_2,...,m_N)$
is generated, where $N$ is assumed to be fixed, and $m_i \in \{0,1
\}$, $i=1,2,...,N$. The bit stream is then sent to the channel
encoder. The encoding process begins by dividing the uncoded message
$\mathcal{M}_N$ into blocks of equal length denoted by
$\mathcal{B}_j \triangleq (m_{_{(j-1)k+1}},...,m_{jk})$,
$j=1,...,\frac{N}{k}$, where $k$ is the length of any particular
$\mathcal{B}_j$, and $N$ is assumed to be divisible by $k$. Each
block $\mathcal{B}_j$ is encoded by a pre-determined channel coding
scheme to generate a coded bit stream $\mathcal{C}_j \triangleq
(a_{_{(j-1)n+1}},...,a_{jn})$, $j=1,...,\frac{N}{k}$, with block
length $n$, where $n$ is either a fixed value (e.g., for block and
convolutional codes) or a random variable (e.g., for LT codes).
\begin{figure}[t]
\centerline{\psfig{figure=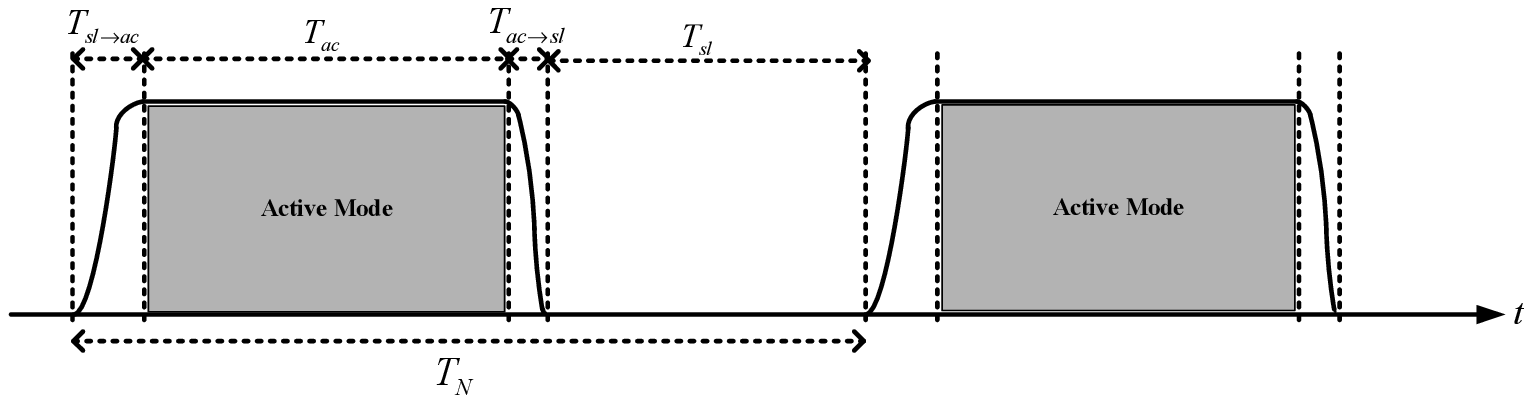,width=3.55in}} \caption{A
practical duty-cycling process in a proactive WSN. } \label{fig:
Time-Basis}
\end{figure}
The coded stream is then modulated by the FSK scheme and transmitted
to the sink node. Finally, the sensor node returns to sleep mode,
and all the circuits are shutdown for sleep mode duration $T_{sl}$.
We denote $T_{tr}$ as the transient mode duration consisting of the
switching time from sleep mode to active mode (i.e., $T_{sl
\rightarrow ac}$) plus the switching time from active mode to sleep
mode (i.e., $T_{ac \rightarrow sl}$), where $T_{ac \rightarrow sl}$
is short enough to be negligible. Under the above considerations,
the sensor/sink nodes have to process one entire $N$-bit message
$\mathcal{M}_N$ during $0 \leq T_{ac} \leq T_N-T_{tr}$, where $T_N
\triangleq T_{tr}+T_{ac}+T_{sl}$ is fixed and $T_{tr} \approx T_{sl
\rightarrow ac}$.

Since sensor nodes in a typical WSN are densely deployed, the
distance between nodes is normally short. Thus, the total circuit
power consumption, defined by $\mathcal{P}_c \triangleq
\mathcal{P}_{ct}+\mathcal{P}_{cr}$, is comparable to the RF transmit
power consumption denoted by $\mathcal{P}_t$, where
$\mathcal{P}_{ct}$ and $\mathcal{P}_{cr}$ represent the circuit
power consumptions for the sensor and sink nodes, respectively.
Taking these into account, the total energy consumption during the
active mode period, denoted by $\mathcal{E}_{ac}$, is given by
$\mathcal{E}_{ac}=(\mathcal{P}_c+\mathcal{P}_t)T_{ac}$. Also,
the power consumption during the
sleep mode duration $T_{sl}$ is much smaller than the power
consumption in the active mode (due to the low sleep mode leakage
current) to be negligible. As a result, we have the following
definition.

\begin{defi}\label{Definition01}
\textbf{(Performance Metric):} The energy efficiency, referred to as
the performance metric of the proposed WSN, can be measured by the
total energy consumption in each period $T_N$ corresponding to
$N$-bit message $\mathcal{M}_N$ as follows:
\begin{equation}\label{total_energy1}
\mathcal{E}_N =
(\mathcal{P}_c+\mathcal{P}_t)T_{ac}+\mathcal{P}_{tr}T_{tr},
\end{equation}
where $\mathcal{P}_{tr}$ is the circuit power consumption during the
transient mode period.
\end{defi}
We use (\ref{total_energy1}) to investigate and compare the energy
efficiency of uncoded and coded FSK for various channel coding
schemes in the subsequent sections.

\textbf{Channel Model:} The choice of low transmission power in WSNs
results in several consequences to the channel model. It is shown by
Friis \cite{Friis1946} that a low transmission power implies a small
range. For short-range transmission scenarios, the root mean square
(rms) delay spread is in the range of nanoseconds
\cite{Karl_Book2005} which is small compared to the symbol duration
$T_s=16M~\mu$s obtained from the bandwidth $B=\frac{M}{T_s}=62.5$ KHz
in the IEEE 802.15.4 standard, where $M$ is the constellation size of M-ary FSK
\cite[pp. 114-115]{Xiong_2006} and \cite[p. 49]{IEEE_802_15_4_2006}.
Thus, it is reasonable to expect a flat-fading channel model for
WSNs. In addition, many transmission environments include
significant obstacle and structural interference by obstacles (such
as wall, doors, furniture, etc), which leads to reduced
Line-Of-Sight (LOS) components. This behavior suggests a Rayleigh
fading channel model. Under the above considerations, the channel
model between the sensor and sink nodes is assumed to be Rayleigh
flat-fading with path-loss. This assumption is used in many works in
the literature (e.g., see \cite{TangITWC0407} for WSNs). For this
model, we assume that the channel is constant during the
transmission of a codeword, but may vary from one codeword to
another. We denote the fading channel coefficient corresponding to
symbol $i$ as $h_{i}$, where the amplitude $\big\vert h_{i}
\big\vert$ is Rayleigh distributed with probability density function
(pdf) $f_{\vert h_{i}
\vert}(r)=\frac{2r}{\Omega}e^{-\frac{r^2}{\Omega}},~r \geq 0$, where
$\Omega \triangleq \mathbb{E}\left[\vert h_{i} \vert^2\right]$
\cite{Proakis2001}.

To model the path-loss of a link where the transmitter and receiver
are separated by distance $d$, let denote $\mathcal{P}_t$ and
$\mathcal{P}_r$ as the transmitted and the received signal powers,
respectively. For a $\eta^{th}$-power path-loss channel, the channel
gain factor is given by $\mathcal{L}_d \triangleq
\frac{\mathcal{P}_t}{\mathcal{P}_r}=M_ld^\eta \mathcal{L}_1$, where
$M_l$ is the gain margin which accounts for the effects of hardware
process variations, background noise and $\mathcal{L}_1 \triangleq
\frac{(4 \pi)^2}{\mathcal{G}_t \mathcal{G}_r \lambda^2}$ is the gain
factor at $d=1$ meter which is specified by the transmitter and
receiver antenna gains $\mathcal{G}_t$ and $\mathcal{G}_r$, and
wavelength $\lambda$ (e.g., \cite{Cui_GoldsmithITWC0905}). As a
result, when both fading and path-loss are considered, the
instantaneous channel coefficient becomes $G_{i} \triangleq
\frac{h_{i}}{\sqrt{\mathcal{L}_d}}$. Denoting $x_i(t)$ as the
transmitted signal with energy $\mathcal{E}_{t}$, the received
signal at the sink node is given by
$y_{i}(t)=G_{i}x_{i}(t)+n_{i}(t)$, where $n_{i}(t)$ is AWGN at the
sink node with two-sided power spectral density given by
$\frac{N_{0}}{2}$. Under the above considerations, the instantaneous
Signal-to-Noise Ratio (SNR) corresponding to symbol $i$ can be
computed as $\gamma_{i}=\frac{\vert G_{i}\vert^2
\mathcal{E}_t}{N_0}$. Under the assumption of a Rayleigh fading
channel model, $\gamma_{i}$ is chi-square distributed with 2 degrees
of freedom and with pdf
$f_{\gamma}(\gamma_{i})=\frac{1}{\bar{\gamma}}e^{
-\frac{\gamma_{i}}{\bar{\gamma}}}$, where $\bar{\gamma} \triangleq
\mathbb{E}[\vert
G_{i}\vert^2]\frac{\mathcal{E}_t}{N_0}=\frac{\Omega}{\mathcal{L}_d}\frac{\mathcal{E}_t}{N_0}$
denotes the average received SNR.

\section{Energy Consumption of Uncoded Scheme}\label{uncoded_MFSK}
In this section, we consider the uncoded M-ary FSK modulation where
$M$ orthogonal carriers can be mapped into $b \triangleq \log_{2}M$
bits. Among various sinusoidal carrier-based modulation techniques,
FSK has been found to provide a good compromise between simple radio
architecture, low-power consumption, and requirements on linearity
of the modulation scheme \cite{JamshidICASSP_2010, Jamshid_IET2010,
Cui_GoldsmithITWC0905}. Also, this scheme is used in some IEEE
standards (e.g., \cite{IEEE_P802_15}). Since we have $b$ bits during
each symbol period $T_{s}$, we can write
\begin{equation}\label{active1}
T_{ac}=\dfrac{N}{b}T_{s}\stackrel{(a)}{=}\dfrac{MN}{B\log_2 M}.
\end{equation}
where $(a)$ comes from the bandwidth $B \approx M\times\Delta f $
with the minimum carrier separation $\Delta f=\frac{1}{T_s}$ for
MFSK with the non-coherent detector \cite[pp. 114-115]{Xiong_2006}. It is
shown in \cite{JamshidICASSP_2010} that the transmit energy
consumption per each symbol for an uncoded MFSK with non-coherent
detector is obtained as
\begin{eqnarray}
\mathcal{E}_t &\triangleq& \mathcal{P}_t T_s \approx \left[\left(
1-(1-P_s)^{\frac{1}{M-1}}\right)^{-1}-2 \right]\dfrac{\mathcal{L}_d
N_0}{\Omega}\\
\notag&\stackrel{(a)}{=}& \left[\left(
1-\left(1-\frac{2(M-1)}{M}P_b\right)^{\frac{1}{M-1}}\right)^{-1}-2
\right]\dfrac{\mathcal{L}_d N_0}{\Omega}\\
\label{deriv1}
\end{eqnarray}
where $(a)$ comes from the fact that the relationship between the
average Symbol Error Rate (SER) $P_s$ and the average Bit Error Rate
(BER) $P_b$ of MFSK is given by $P_s=\frac{2(M-1)}{M}P_b$ \cite[p.
262]{Proakis2001}. As a result, the output energy consumption of
transmitting $N$-bit during $T_{ac}$ of an uncoded MFSK is computed
from (\ref{active1}) as follows:
\begin{eqnarray}
\notag \mathcal{P}_t T_{ac}&=& \dfrac{T_{ac}}{T_s}\mathcal{E}_t
\approx \left[\left(
1-\left(1-\frac{2(M-1)}{M}P_b\right)^{\frac{1}{M-1}}\right)^{-1}-2
\right]\\
\label{energy_trans1}&\times&\dfrac{\mathcal{L}_d N_0}{\Omega}
\dfrac{N}{\log_2 M}.
\end{eqnarray}
For the sensor node with uncoded MFSK, we denote the power
consumption of frequency synthesizer, filters and power amplifier as
$\mathcal{P}_{Sy}$, $\mathcal{P}_{Filt}$ and $\mathcal{P}_{Amp}$,
respectively. In this case, the circuit power consumption of the
sensor node with uncoded MFSK can be obtained as
\begin{equation}\label{sensor_power}
\mathcal{P}_{ct}=\mathcal{P}_{Sy}+\mathcal{P}_{Filt}+\mathcal{P}_{Amp},
\end{equation}
where $\mathcal{P}_{Amp}=\alpha \mathcal{P}_{t}$ with $\alpha=0.33$
\cite{Cui_GoldsmithITWC0905}, \cite{TangITWC0407}. In addition, the
power consumption of the sink circuitry with uncoded MFSK scheme can
be obtained as
\begin{equation}\label{sink_power}
\mathcal{P}_{cr}=\mathcal{P}_{LNA}+M\times(\mathcal{P}_{Filr}+\mathcal{P}_{ED})+\mathcal{P}_{IFA}+\mathcal{P}_{ADC},
\end{equation}
where $\mathcal{P}_{LNA}$, $\mathcal{P}_{Filr}$, $\mathcal{P}_{ED}$,
$\mathcal{P}_{IFA}$ and $\mathcal{P}_{ADC}$ denote the power
consumption of Low-Noise Amplifier (LNA), filters, envelop detector,
IF amplifier and ADC, respectively \cite{JamshidICASSP_2010}. Since,
the power consumption during transient mode period $T_{tr}$ is
governed by the frequency synthesizer in both transmitter and the
receiver \cite{Wang_ISLPED2001}, the energy consumption during
$T_{tr}$ is obtained as $\mathcal{P}_{tr}T_{tr}=2
\mathcal{P}_{Sy}T_{tr}$ \cite{Cui_GoldsmithITWC0905}. Substituting
(\ref{active1}) and (\ref{energy_trans1}) in (\ref{total_energy1}),
the total energy consumption of an uncoded MFSK for transmitting
$N$-bit information in each period $T_N$ for a given $P_b$ is
obtained as
\begin{eqnarray}
\notag \mathcal{E}_N &=& (1+\alpha) \left[\left(
1-\left(1-\dfrac{2(M-1)}{M}P_b\right)^{\frac{1}{M-1}}\right)^{-1}-2
\right]\\
\notag&\times&\dfrac{\mathcal{L}_d N_0}{\Omega} \dfrac{N}{\log_2 M}+
(\mathcal{P}_{c}-\mathcal{P}_{Amp})\dfrac{MN}{B\log_{2}M}+\\
\label{energy_totFSK}&&2 \mathcal{P}_{Sy}T_{tr}.
\end{eqnarray}

For energy optimal designs, the impact of channel coding on the
energy efficiency of the proposed WSN must be considered as well. It
is a well known fact that channel coding is a classical approach
used to improve the link reliability along with the transmitter
energy saving due to providing the coding gain \cite{Proakis2001}.
However, the energy saving comes at the cost of extra energy spent
in transmitting the redundant bits in codewords as well as the
additional energy consumption in the process of encoding/decoding.
For a specific transmission distance $d$, if these extra energy
consumptions outweigh the transmit energy saving due to the channel
coding, the coded system would not be energy-efficient compared with
an uncoded system. In the subsequent sections, we will argue the
above problem and determine at what distance use of specific channel
coding becomes energy-efficient compared to uncoded systems. In
particular, we will show in Section \ref{simulation_Ch5} that the LT
coded modulation surpasses this distance constraint in the proposed
WSN.

\section{Energy Consumption Analysis of LT Coded Modulation}\label{analysis_Ch4}
In this section, we present the first in-depth analysis on the
energy efficiency of LT coded modulation for the proposed proactive
WSN. To get more insight into how channel coding affects the circuit
and RF signal energy consumptions in the system, we modify the
energy concepts in Section \ref{uncoded_MFSK}, in particular, the
total energy consumption expression in (\ref{energy_totFSK}) based
on the coding gain, code rate and the computation energy. We further
present the first study on the tradeoff between LT code rate and
coding gain required to achieve a certain BER, and the effect of
this tradeoff on the total energy consumption of LT coded MFSK for
different transmission distances.

\subsection{Energy Efficiency of Coded System}
For an arbitrary channel coding scheme, each $k$-bit message
$\mathcal{B}_j \in \mathcal{M}_N$ is encoded into the codeword
$\mathcal{C}_j$ with block length $n$ and code rate $R_c \triangleq
\frac{k}{n}$. In this case, the number of transmitted bits in $T_N$
is increased from $N$-bit uncoded message to
$\frac{N}{R_c}=\frac{N}{k}n$ bits coded one. To compute the energy
consumption of coded scheme, we use the fact that channel coding
reduces the required average SNR value to achieve a given BER (i.e.,
the same BER as uncoded one). Taking this into account, the proposed
WSN benefits in transmission energy saving of coded modulation
specified by
$\mathcal{E}_{t,c}=\frac{\mathcal{E}_{t}}{\Upsilon_{c}}$, where
$\Upsilon_{c} \geq 1$ is the coding gain\footnote{Denoting
$\bar{\gamma}
=\frac{\Omega}{\mathcal{L}_d}\frac{\mathcal{E}_t}{N_0}$ and
$\bar{\gamma}_{c}
=\frac{\Omega}{\mathcal{L}_d}\frac{\mathcal{E}_{t,c}}{N_0}$ as the
average SNR of uncoded and coded schemes, respectively, the
\emph{coding gain} (expressed in dB) is defined as the difference
between the values of $\bar{\gamma}$ and $\bar{\gamma}_{c}$ required
to achieve a certain BER, where
$\mathcal{E}_{t,c}=\frac{\mathcal{E}_{t}}{\Upsilon_{c}}$.} of the
utilized coded MFSK. It should be noted that the cost of this energy
saving is the bandwidth expansion $\frac{B}{R_{c}}$. In order to
keep the bandwidth of the coded system the same as that of the
uncoded case, we must keep the information transmission rate
constant, i.e., the symbol duration $T_{s}$ of uncoded and coded
MFSK would be the same. However, the active mode duration increases
from $T_{ac}=\frac{N}{b}T_{s}$ in the uncoded system to
\begin{equation}\label{BCH_active}
T_{ac,c}=\frac{N}{bR_{c}}T_{s}=\frac{T_{ac}}{R_{c}}
\end{equation}
for the coded case. Thus, one would assume that the total time $T_N$
increases to $\frac{T_N}{R_c}$ for the coded scenario. It is worth
mentioning that the active mode period in the coded case is upper
bounded by $\frac{T_N}{R_c}-T_{tr}$. As a result, the maximum
constellation size $M$, denoted by $M_{max}\triangleq 2^{b_{max}}$,
for the coded MFSK is calculated by
$\frac{2^{b_{max}}}{b_{max}}=\frac{B
R_{c}}{N}(\frac{T_{N}}{R_{c}}-T_{tr})$, which is approximately the
same as that of the uncoded case.

\newcounter{mytempeqncnt}
\begin{figure*}[!t]
\normalsize \setcounter{mytempeqncnt}{\value{equation}}
\setcounter{equation}{10}
\begin{eqnarray}
\left(\left[1-\left(1-\frac{2(M-1)}{M}P_b\right)^{\frac{1}{M-1}}\right]^{-1}-2\right) \dfrac{1}{\log_2 M}&=& \left(\left[1-e^{\frac{1}{M-1}\ln(1-\frac{2(M-1)}{M}P_b)}\right]^{-1}-2\right)\dfrac{1}{\log_2 M}\\
&\stackrel{(a)}{\approx}& \left(\left[1-e^{-\frac{2P_b}{M}}\right]^{-1}-2\right)\dfrac{1}{\log_2 M}\\
\label{mono10}&\stackrel{(b)}{\approx}&\left(\dfrac{M}{2P_b}-2\right)\dfrac{1}{\log_2
M},
\end{eqnarray}
\setcounter{equation}{\value{mytempeqncnt}} \hrulefill \vspace*{4pt}
\end{figure*}

We denote $\mathcal{E}_{enc}$ and $\mathcal{E}_{dec}$ as the
computation energy of the encoder and decoder for each information
bit, respectively. Thus, the total computation energy cost of the
coding components for $\frac{N}{R_{c}}$ bits is obtained as
$N\frac{\mathcal{E}_{enc}+\mathcal{E}_{dec}}{R_{c}}$. Substituting
(\ref{deriv1}) in
$\mathcal{E}_{t,c}=\frac{\mathcal{E}_{t}}{\Upsilon_{c}}$, and using
(\ref{sensor_power}), (\ref{sink_power}) and (\ref{BCH_active}), the
total energy consumption of transmitting $\frac{N}{R_c}$ bits in
each period $\frac{T_N}{R_C}$ for an arbitrarily coded MFSK, and a
given $P_b$ is obtained as
\begin{eqnarray}
\notag \mathcal{E}_{N,c} &=& (1+\alpha) \left[\left(
1-\left(1-\dfrac{2(M-1)}{M}P_b\right)^{\frac{1}{M-1}}\right)^{-1}-2
\right]\\
\notag&\times&\dfrac{\mathcal{L}_d N_0}{\Omega \Upsilon_{c}}
\dfrac{N}{R_{c}\log_2 M}+
(\mathcal{P}_{c}-\mathcal{P}_{Amp})\dfrac{MN}{BR_{c} \log_{2}M}+ \\
\label{energy_totalcodedFSK} && 2
\mathcal{P}_{Sy}T_{tr}+N\frac{\mathcal{E}_{enc}+\mathcal{E}_{dec}}{R_{c}}.
\end{eqnarray}
To make a fair comparison between the uncoded and coded modulation,
we use the same BER and drop the subscript ``$c$'' for $P_b$ in the
coded case. Thus, the optimization goal is to minimize the objective
function $\mathcal{E}_{N,c}$ over modulation and coding parameters.
This is achieved by finding the optimum constellation size $M$ under
the constraint $2\leq M \leq M_{max}$ for a specific channel coding
scheme, and then minimize $\mathcal{E}_{N,c}$ over the coding
parameters.

For the above optimization problem, we consider two scenarios: $i)$
fixed-rate codes (e.g., BCH and convolutional codes), and $ii)$
variable rate codes (e.g., LT codes). To find the optimum $M$ for a
given fixed-rate code, we prove that (\ref{energy_totalcodedFSK}) is
a monotonically increasing function of $M$ for every value of $d$
and $\eta$. Since $B$ and $N$ are fixed and $R_c$ is independent of
$M$, it is concluded that the second term in
(\ref{energy_totalcodedFSK}) is a monotonically increasing function
of $M$. Also, from the first term in (\ref{energy_totalcodedFSK}),
we have (11)-(13) in the top of this page,
where $(a)$ comes from the approximation $\ln(1-z)\approx -z,~~|z|
\ll 1$, and the fact that $P_b$ scales as $o(1)$. Also, $(b)$
follows from the approximation
$e^{-z}=\sum_{n=0}^{\infty}(-1)^{n}\frac{z^n}{n!}\approx
1-z,~~|z|\ll1$. On the other hand, it is shown in \cite{Proakis2001}
that $\Upsilon_c$ for MFSK is a decreasing function of $M$. Thus, it
is concluded from (\ref{mono10}) that the first term in
(\ref{energy_totalcodedFSK}) is also a monotonically increasing
function of $M$. As a result, the minimum total energy consumption
$\mathcal{E}_{N,c}$ for a given fixed-rate code is achieved at
$\hat{M}=2$.

In the next section, we evaluate the above optimization problem for
the LT codes using some simulation studies on the probability mass
function of the LT code rate and coding gain. We show that the LT
code parameters depend strongly on the constellation size $M$ and
exhibit different trends over fixed-rate codes. In addition, we
present some beneficial uses of LT codes over block and
convolutional codes in managing the energy consumption for different
channel realizations.

\subsection{Energy Optimality of LT Codes}
LT codes are the first class of Fountain codes which usually
specified by the number of input bits $k$ and the output-node degree
distribution $\mathcal{O}(x)$. Without loss of generality and for
ease of our analysis, we assume that a single $k$-bit message
$\mathcal{B}_1 \triangleq (m_1,m_2,...,m_k)\in \mathcal{M}_N$ is
encoded to codeword $\mathcal{C}_1 \triangleq (a_1,a_2,...,a_n)$.
Each single coded bit $a_i$ is generated based on the encoding
protocol proposed in \cite{LubyFOCS2002}: $i)$ randomly choose a
degree $1 \leq \mathcal{D} \leq k$ from $\mathcal{O}(x)$, $ii)$
using a uniform distribution, randomly choose $\mathcal{D}$ distinct
input bits, and calculate the encoded bit $a_i$ as the XOR-sum of
these $\mathcal{D}$ bits. The above encoding process defines a
\emph{bipartite graph} connecting encoded nodes to input nodes (see,
e.g., Fig. \ref{fig: LT Encoder}). It is seen that the LT encoding
process is extremely simple and has very low energy consumption.
Unlike block and convolutional codes, in which the codeword block
length is fixed, for the above LT code, $n$ is a variable parameter,
resulting in a random variable LT code rate $R^{LT}_c \triangleq
\frac{k}{n}$. More precisely, $a_n \in \mathcal{C}_1$ is the last
bit generated at the output of LT encoder before receiving the
acknowledgement signal from the sink node indicating termination of
a successful decoding process. This inherent property of LT codes
means they can vary their codeword block lengths to adapt to any
wireless channel condition between the sensor and the sink nodes.

\begin{figure}[t]
\centerline{\psfig{figure=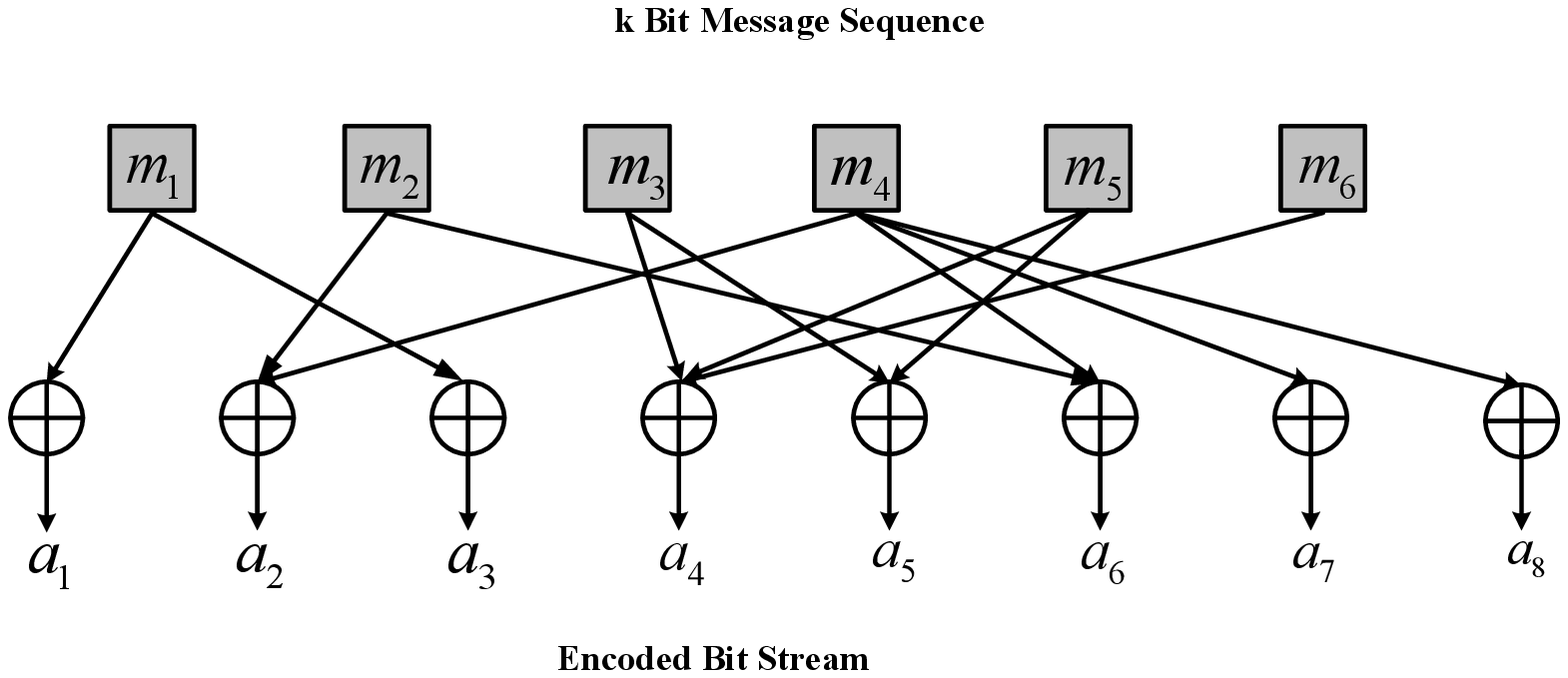,width=3.25in}} \caption{
Bipartite graph of an LT code with $k=6$ and $n=8$. } \label{fig: LT
Encoder}
\end{figure}

To describe the output-node degree distribution used in this work,
let $\mu_i$, $i=1,...,k$, denote the probability that an output node
has degree $i$. Following the notation of
\cite{Shokrollahi_ITIT0606}, the output-node degree distribution of
an LT code has the polynomial form $\mathcal{O}(x) \triangleq
\sum_{i=1}^{k}\mu_i x^i$ with the property that $\mathcal{O}(1) =
\sum_{i=1}^{k}\mu_i =1$. Typically, optimizing the output-node
degree distribution for a specific wireless channel model is a
crucial task in designing LT codes. In fact, for wireless fading
channels, it is still an open problem, what the ``optimal"
$\mathcal{O}(x)$ is. In this work, we use the following output-node
degree distribution which was optimized for a BSC using a
hard-decision decoder \cite{David_Thesis2008}:
\begin{eqnarray}
\notag \mathcal{O}(x)&=&0.00466x+0.55545x^2+0.09743x^3+\\
\notag &&0.17506x^5+0.03774x^8+0.08202x^{14}+\\
\label{degree}&&0.01775x^{33}+0.02989x^{100}.
\end{eqnarray}

The LT decoder at the sink node can recover the original $k$-bit
message $\mathcal{B}_1$ with high probability after receiving any
$(1+\epsilon)k$ bits in its buffer, where $\epsilon$ depends upon
the LT code design \cite{Shokrollahi_ITIT0606}. For this recovery
process, the LT decoder needs to correctly reconstruct the bipartite
graph of an LT code. One practical approach suitable for the
proposed WSN model is that the LT encoder and decoder use identical
pseudo-random generators with a common seed value which may reduce
the complexity further. In this work, we assume that the sink node
recovers $k$-bit message $\mathcal{B}_1$ using a simple
hard-decision ``\emph{ternary message passing}" decoder in a nearly
identical manner to the ``\emph{Algorithm E}'' decoder in
\cite{RichardsonITIT0201} for Low-Density Parity-Check (LDPC)
codes\footnote{Description of the \emph{ternary message passing}
decoding is out of scope of this work, and the reader is referred to
Chapter 4 in \cite{David_Thesis2008} for more details.}. Also, the
degree distribution $\mathcal{O}(x)$ in (\ref{degree}) was optimized
for a ternary decoder in a BSC and we are aware of no better
$\mathcal{O}(x)$ for the ternary decoder in Rayleigh fading
channels.

Unlike fixed-rate codes in which the active mode duration of coded
MFSK is fixed, for the LT coded MFSK, we have a non-fixed value for
\begin{equation}\label{active_LT}
T_{ac,c}^{LT}=\dfrac{N}{b R_{c}^{LT}}T_s=\dfrac{MN}{BR^{LT}_{c}
\log_{2}M}.
\end{equation}
An interesting point raised from (\ref{active_LT}) is that
$T_{ac,c}^{LT}$ is a function of the random variable $R_c^{LT}$
which results in an inherent adaptive duty-cycling for power
management in each channel condition without any channel state
information fed back from the sink node to the sensor node.
Recalling from (\ref{energy_totalcodedFSK}), we have
\begin{eqnarray}
\notag \mathcal{E}^{LT}_{N,c} &=& (1+\alpha) \left[\left(
1-\left(1-\dfrac{2(M-1)}{M}P_b\right)^{\frac{1}{M-1}}\right)^{-1}-2
\right]\\
\notag &\times&\dfrac{\mathcal{L}_d N_0}{\Omega \Upsilon^{LT}_{c}}
\dfrac{N}{R^{LT}_{c}\log_2 M}+
(\mathcal{P}_{c}-\mathcal{P}_{Amp})\dfrac{MN}{BR^{LT}_{c}
\log_{2}M}+ \\
\notag&&2
\mathcal{P}_{Sy}T_{tr}+N\frac{\mathcal{E}^{LT}_{enc}+\mathcal{E}^{LT}_{dec}}{R^{LT}_{c}}.
\end{eqnarray}
where the main goal is to minimize $\mathcal{E}^{LT}_{N,c}$ over $M$
and the coding parameters. Toward this goal, we first compute the LT
code rate and the corresponding LT coding gain. Let us begin with
the case of asymptotic LT code rate, where the number of input bits
$k$ goes to infinity. It is shown that the LT code rate is obtained
asymptotically as a fixed value of $R^{LT}_c \approx
\frac{\mathcal{O}_{ave}}{\mathcal{I}_{ave}}$ for large values of
$k$, where $\mathcal{O}_{ave}$ and $\mathcal{I}_{ave}$ represent the
average degree of the output and the input nodes in the bipartite
graph, respectively (see Appendix I for the proof).

When using finite-$k$ LT codes, we are treating the channel as
static over one block length. In this case, the rate of the LT code
for any block can be chosen to achieve the desired performance for
that block. More precisely, for the particular block we are
concerned with, the receiver could evaluate the channel
instantaneous SNR, and then determine how many code bits it needs to
collect in order to achieve its given BER target, thus essentially
dynamically selecting its rate.  For the next block, the receiver
would once again evaluate the (new) instantaneous channel SNR and
adjust its rate accordingly to collect more or fewer bits.

It should be noted that there is no (currently known) explicit equation
governing the relationship between the instantaneous SNR and the required
number of decoding bits for an LT code. In this work, we determine
the necessary number of decoding bits in each case through simulation,
as described below:
\begin{itemize}
\item [$\textbf{i)}$] We decide upon a ``\emph{target BER}'' for the decoded bits - e.g., the
decoded message needed an average BER of $10^{-4}$ or better.
\item [$\textbf{ii)}$] Based on the assumption that instantaneous SNR is constant over at
least one block we perform the following steps for a large range of possible
instantaneous SNR:

- Determine through numerical simulation the decoded BER using the LT code rate $R^{LT}_c= 1$.

- If the decoded BER is greater than the target BER, then we reduce the LT code rate $R^{LT}_c$ and try again.

- Repeat the above step until the decoded BER is less than the target BER.

\item [$\textbf{iii)}$] At this point, for the given LT degree distribution, decoder, and SNR
we can identify the highest LT code rate that will yield the target BER
or better.
\end{itemize}
Armed with this information (computed ahead of time), a receiver can determine
the appropriate rate at which to operate and hence determine how many bits it
must collect for a given instantaneous SNR.
This is also based on the common assumption that the receiver is capable of
estimating the instantaneous SNR.

\begin{figure}[t]
\centerline{\psfig{figure=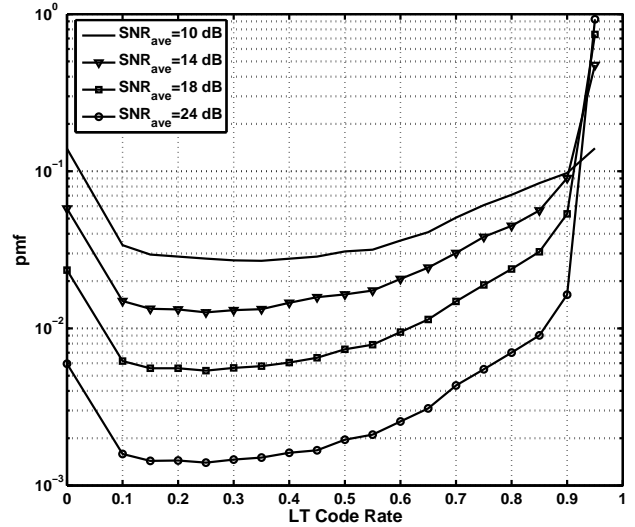,width=3.75in}}
\caption{The pmf of LT code rate for various average SNR and M=2. }
\label{fig: LT code_rate_mpf}
\end{figure}

\begin{figure}[t]
\centerline{\psfig{figure=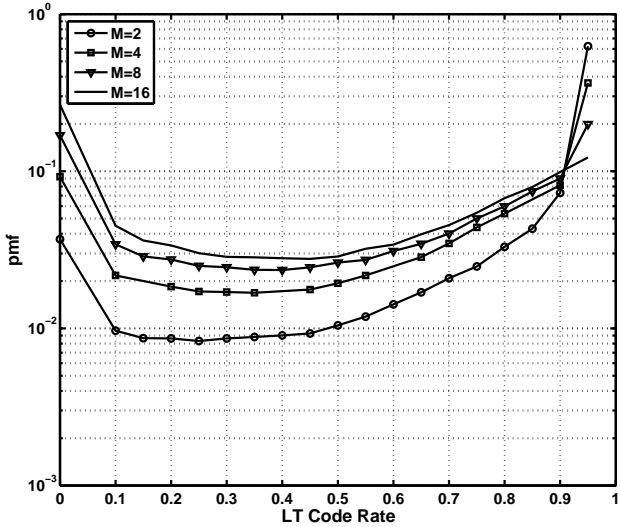,width=3.75in}}
\caption{The pmf of LT code rate for various constellation size $M$
and average SNR=16 dB. } \label{fig: LT code_rate_pmf_M}
\end{figure}

Based on the above arguments and for any given average SNR, the LT code rate is described
by either a probability mass function (pmf) or a probability density
function (pdf) denoted by $P_R(\ell)$. Because it is difficult to
get a closed-form expression of $P_R(\ell)$, we use a discretized
numerical method to calculate the pmf $P_R(\ell)\triangleq
\textrm{Pr}\{R_c^{LT}=\ell \}$, $0 \leq \ell \leq 1$, for different
values of $M$.
We plot the pmfs of the LT code rate in Fig. \ref{fig: LT code_rate_mpf}
for $M=2$ and various average SNR \footnote{It should be noted that
the average SNR is expressed in terms of $\frac{\mathcal{E}_b}{N_0}$,
where $\mathcal{E}_b$ represents the transmit energy per bit.} over a Rayleigh fading channel model.
It is observed that for lower average SNRs the pmfs are larger in
the lower rate regimes (i.e., the pmfs spend more time in the low
rate region). Also, all pmfs exhibit quite a spike for the highest
rate, which makes sense since once the instantaneous SNR hits a
certain critical value, the codes will always decode with a high
rate. Also, Fig. \ref{fig: LT code_rate_pmf_M} illustrates the pmf
of LT code rates for various constellation size $M$ and for average
SNR equal to 16 dB. It can be seen that as $M$ increases, the rate
of the LT code tends to have a pmf with larger values in the lower
rate regions.

\begin{figure*}[t]
\centerline{\psfig{figure=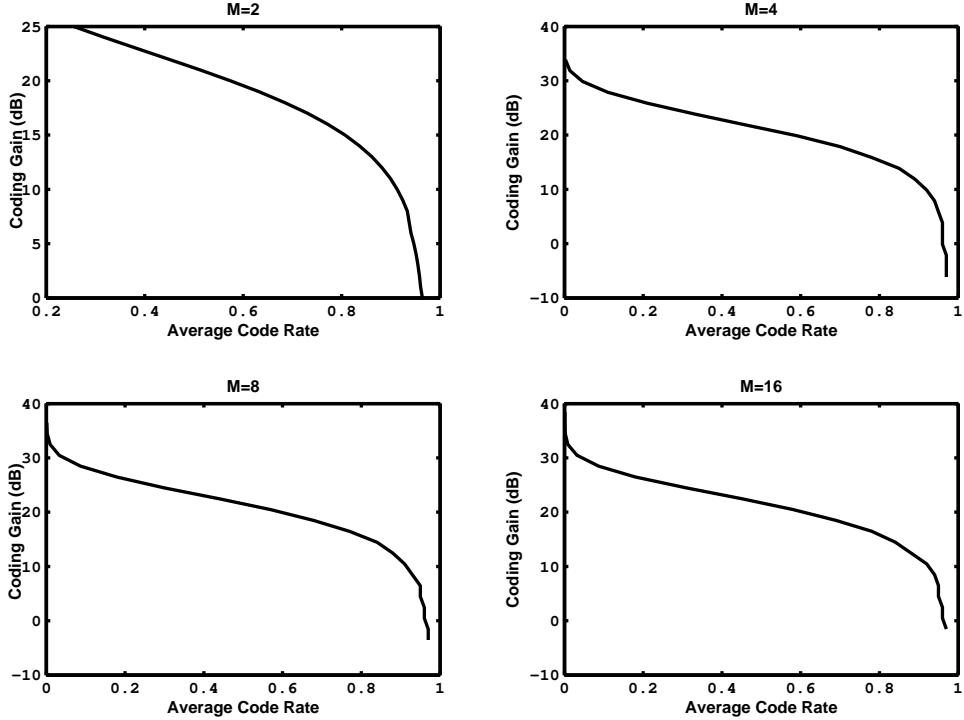,width=6.15in}}
\caption{LT coding gain versus average LT code rate for $P_b=10^{-3}$ and
M=2,4,8,16. } \label{fig: LT code_rate_coding_gain}
\end{figure*}

Table II illustrates the average LT code rates and the
corresponding coding gains of LT coded MFSK using $\mathcal{O}(x)$
in (\ref{degree}), for $M=2,4,8,16$ and given $P_b=10^{-3}$. The
average rate for a certain average SNR is obtained by integrating
the pmf over the rates from $0$ to $1$. To get more insight into
the relationship between the LT code rate and coding gain, we plot
the results of Table II in Fig. \ref{fig: LT code_rate_coding_gain}. It is observed that the LT
code is able to provide a huge coding gain $\Upsilon_{c}^{LT}$ given
$P_b=10^{-3}$, but this gain comes at the expense of a very low
average code rate, which means many additional code bits need to be
sent. This results in higher energy consumption per information bit.
In contrast to the fixed-rate codes in which the coding gain
decreases when $M$ grows, the LT coding gains display different
trends in terms of $M$ as illustrated in Table II. Thus, in
contrast with fixed-rate codes, $\mathcal{E}_{N,c}^{LT}$ is not
necessarily a monotonically increasing function of $M$. In the next
section, we evaluate numerically $\mathcal{E}_{N,c}^{LT}$ in terms
of the optimized modulation and coding parameters compared to the
uncoded and the fixed-rate codes.

An interesting point extracted from Table II is the flexibility of
the LT code to adjust its rate (and its corresponding coding gain)
to suit instantaneous channel conditions in WSNs. For instance in
the case of favorable channel conditions, the LT coded MFSK is able
to achieve $R_{c}^{LT}\approx1$ with $\Upsilon_{c}^{LT}\approx 0$
dB, which is similar to the case of uncoded MFSK, i.e., $n=k$. The
effect of LT code rate flexibility on the total energy consumption
is also observed in the simulation results in the subsequent
section.

\begin{table*}
\label{table0021} \caption{Average LT code rate and Coding gain of
LT coded MFSK over Rayleigh fading model for $P_b=10^{-3}$ and
M=2,4,8,16.} \centering
  \begin{tabular}{|c|cc||c|cc|cc|cc|}
  \hline
    & M=2 &   & &  M=4 &  & M=8 &  & M=16 & \\
   \hline
   $\frac{\mathcal{E}_b}{N_0}$  & Average   &  Coding    & $\frac{\mathcal{E}_b}{N_0}$  & Average   & Coding    & Average   &  Coding    & Average    & Coding    \\
   (dB) & Code Rate &  Gain (dB) & (dB) & Code Rate & Gain (dB) & Code Rate &  Gain (dB) &  Code Rate & Gain (dB) \\
   \hline
   5 & 0.2560 & 25 &  0   & 0.0028 & 33.87 &  0.0012 & 36.46 &  0.0012 & 38.48  \\
   6 & 0.3174 & 24 &  2   & 0.0140 & 31.87 &  0.0024 & 34.46 &  0.0012 & 36.48  \\
   7 & 0.3819 & 23 &  4   & 0.0460 & 29.87 &  0.0095 & 32.46 &  0.0021 & 34.48  \\
   8 & 0.4475 & 22 &  6   & 0.1100 & 27.87 &  0.0330 & 30.46 &  0.0086 & 32.48  \\
   9 & 0.5120 & 21 &  8   & 0.2100 & 25.87 &  0.0870 & 28.46 &  0.0320 & 30.48  \\
   10& 0.5738 & 20 & 10   & 0.3300 & 23.87 &  0.1800 & 26.46 &  0.0870 & 28.48  \\
   11& 0.6315 & 19 & 12   & 0.4600 & 21.87 &  0.3000 & 24.46 &  0.1800 & 26.48  \\
   12& 0.6840 & 18 & 14   & 0.5900 & 19.87 &  0.4400 & 22.46 &  0.3100 & 24.48  \\
   13& 0.7307 & 17 & 16   & 0.7000 & 17.87 &  0.5700 & 20.46 &  0.4500 & 22.48  \\
   14& 0.7716 & 16 & 18   & 0.7800 & 15.87 &  0.6800 & 18.46 &  0.5800 & 20.48  \\
   15& 0.8067 & 15 & 20   & 0.8500 & 13.87 &  0.7700 & 16.46 &  0.6900 & 18.48  \\
   16& 0.8365 & 14 & 22   & 0.8900 & 11.87 &  0.8400 & 14.46 &  0.7800 & 16.48  \\
   17& 0.8614 & 13 & 24   & 0.9200 & 9.87  &  0.8800 & 12.46 &  0.8400 & 14.48  \\
   18& 0.8821 & 12 & 26   & 0.9400 & 7.87  &  0.9100 & 10.46 &  0.8800 & 12.48  \\
   19& 0.8991 & 11 & 28   & 0.9500 & 5.87  &  0.9300 & 8.46  &  0.9200 & 10.48  \\
   20& 0.9130 & 10 & 30   & 0.9600 & 3.87  &  0.9500 & 6.46  &  0.9400 & 8.48   \\
   22& 0.9333 & 8  & 32   & 0.9600 & 1.87  &  0.9500 & 4.46  &  0.9500 & 6.48   \\
   24& 0.9466 & 6  & 34   & 0.9600 & -0.13 &  0.9600 & 2.46  &  0.9500 & 4.48   \\
   26& 0.9551 & 4  & 36   & 0.9700 & -2.13 &  0.9600 & 0.46  &  0.9600 & 2.48   \\
   28& 0.9606 & 2  & 38   & 0.9700 & -4.13 &  0.9700 & -1.54 &  0.9600 & 0.48   \\
   30& 0.9640 & 0  & 40   & 0.9700 & -6.13 &  0.9700 & -3.54 &  0.9700 & -1.52  \\

   \hline
   \end{tabular}
\end{table*}

\section{Numerical Evaluations}\label{simulation_Ch5}
In this section, we present some numerical evaluations using
realistic parameters from the IEEE 802.15.4 standard and
state-of-the art technology to confirm the energy efficiency
analysis of uncoded and coded MFSK modulations discussed in Sections
\ref{uncoded_MFSK} and \ref{analysis_Ch4}.

\subsection{Experimental Setup}
We assume that MFSK operates in the carrier frequency $f_0=$2.4 GHz
Industrial Scientist and Medical (ISM) unlicensed band utilized in
the IEEE 802.15.14 standard \cite{IEEE_802_15_4_2006}. According to
the FCC 15.247 RSS-210 standard for United States/Canada, the
maximum allowed antenna gain is 6 dBi \cite{FreeScale2007}. In this
work, we assume that $\mathcal{G}_t=\mathcal{G}_r=5$ dBi. Thus for
the $f_0=$2.4 GHz, $\mathcal{L}_1~ \textrm{(dB)}\triangleq
10\log_{10}\left(\frac{(4 \pi)^2}{\mathcal{G}_t \mathcal{G}_r
\lambda^2}\right) \approx 30~ \textrm{dB}$, where $\lambda
\triangleq \frac{3\times 10^8}{f_0}=0.125$ m. We assume that in each
period $T_N$, the data frame $N=1024$ bytes (or equivalently
$N=8192$ bits) is generated, where $T_N$ is assumed to be 1.4 s. The
channel bandwidth is set to the $B=62.5$ KHz, according to IEEE
802.15.4 \cite[p. 49]{IEEE_802_15_4_2006}. It is concluded from
$\frac{2^{b_{max}}}{b_{max}}=\frac{BR_c}{N}(\frac{T_{N}}{R_c}-T_{tr})\approx
\frac{B}{N}T_N$ that $M_{max} \approx 64$ (or equivalently $b_{max}
\approx 6$) for MFSK. Table III summarizes the system parameters for
simulation\footnote{To make a fair comparison between the energy
consumption of different communication
schemes, the bandwidth and the BER are assumed to be the same for all the schemes.} \cite{Cui_GoldsmithITWC0905, TangITWC0407,
Bevilacqua_IJSSC1204, Wang_ISLPED2001}. The results in Tables II
are also used to compare the energy efficiency of uncoded and
coded MFSK schemes.

\begin{table}
\label{table001} \caption{System Evaluation Parameters} \centering
  \begin{tabular}{lll}
  \hline
   $B=62.5$ KHz          & $N_0=-180$ dB               & $\mathcal{P}_{ADC}=7$ mw \\

   $M_l=40$ dB           & $\mathcal{P}_{Sy}=10$ mw    &  $\mathcal{P}_{LNA}=9$ mw\\

   $\mathcal{L}_1=30$ dB & $\mathcal{P}_{Filt}=2.5$ mw & $\mathcal{P}_{ED}=3$ mw  \\

   $\eta=3.5$            & $\mathcal{P}_{Filr}=2.5$ mw & $\mathcal{P}_{IFA}=3$ mw \\

   $\Omega=1$            &  $T_N=1.4$ sec              & $T_{tr}=5~\mu s$      \\

  \hline
  \end{tabular}
\end{table}

In order to estimate the computation energy of a specific channel
coding, we use the ARM7TDMI core which is the industry's most widely
used 32-bit embedded RISC microprocessor for an accurate power
simulation \cite{ARMTech2004}. For the energy and power
calculations, the relations
$\mathcal{E}_{operation}=\dfrac{n_c}{n_o}\times \mathcal{E}_{Hz}$
and $\mathcal{E}_{block}=n_o \times \mathcal{E}_{operation}$ are
used, where
\begin{itemize}
\item [] $n_o:$ number of operations per block.
\item [] $n_c:$ number of clock cycles on $n_o$ operations.
\item [] $\mathcal{E}_{Hz}:$ total dynamic power consumption per
Hertz (W/Hz) of a single calculation cycle.
\end{itemize}
In addition, for the purpose of comparative evaluation, we use some
classical BCH$(n,k,t)$ codes with $t$-error correction capability,
and convolutional codes $(n,k,L)$ with the constraint length $L$.
These codes are widely utilized in IEEE standards \cite{IEEE802.15,
IEEE802.16}. We use hard-decision prior to decoding for the BCH and
convolutional codes. The main reason for using the hard-decision
here is to make a fair comparison to the LT code, since this code
involved a hard-decision in ternary decoder.

\subsection{Optimal Configuration}
\begin{figure}[t]
\centerline{\psfig{figure=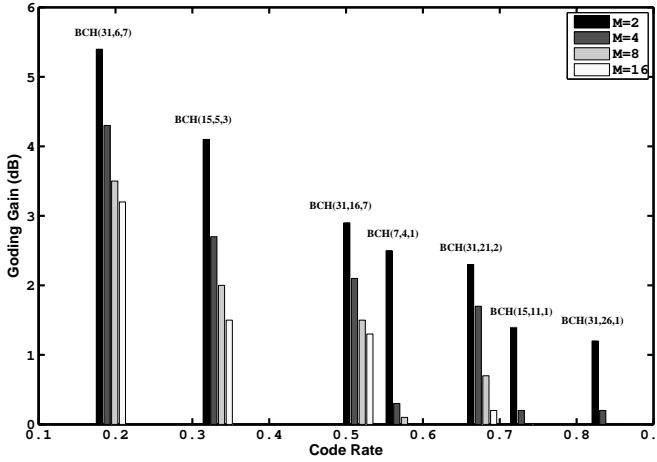,width=4.15in}}
\caption{Coding gain versus code rate of some practical BCH$(n,k,t)$ codes for $P_b=10^{-3}$ and
M=2,4,8,16. } \label{fig: BCH code_rate_coding_gain}
\end{figure}

As a starting point, we obtain the coding gain of some practical BCH
and convolutional codes with MFSK for different constellation size
$M$ and given $P_b=(10^{-3},10^{-4})$ in Table IV. We observe from
Table IV that there is a tradeoff between coding gain and decoder
complexity for the BCH codes. In fact, achieving a higher coding
gain for a given $M$, requires a more complex decoding process,
(i.e., higher $t$) with more circuit power consumption. In addition,
it is seen from Fig. \ref{fig: BCH code_rate_coding_gain} that the coding gain of the BCH code is a monotonically
decreasing function of $M$ as expected. For a given $M$, the
convolutional codes with lower rates and higher constraint lengths
achieve greater coding gains. In contrast to
\cite{Cui_GoldsmithITWC0905}, where the authors assume a fixed
convolutional coding gain for every value of $M$, it is observed
that the coding gain of convolutional coded MFSK is a monotonically
decreasing function of $M$. By comparing the results in Table II
with those in Table IV for BCH and convolutional codes, one observes
that LT codes outperform the other coding schemes in energy saving
at comparable rates.

\begin{table*}
\label{table01} \caption{Coding gain ($dB$) of BCH and convolutional
coded MFSK over Rayleigh fading for BER$=(10^{-3},10^{-4})$.}
\centering
  \begin{tabular}{|l|c| c c c c c c|}
  \hline
  BCH Code$(n,k,t)$  & $R^{BC}_c$ & M=2  & M=4  & M=8  &  M=16  &  M=32  &  M=64
  \\
  \hline
   BCH $(7,4,1)$   & 0.571 &   $(2.5,2.8)$   &   $(0.3,0.4)$    &    $(0.1,0.2)$  &  $(0.0,0.0)$   &  $(0.0,0.0)$   &  $(0.0,0.0)$   \\
  \hline
   BCH $(15,11,1)$ & 0.733 &   $(1.4,1.6)$   &   $(0.2,0.3)$    &    $(0.0,0.0)$  &  $(0.0,0.0)$   &  $(0.0,0.0)$   &  $(0.0,0.0)$   \\
   \hline
   BCH $(15,7,2)$  & 0.467 &  $(2.4,3.3)$   &   $(2.0,2.3)$    &    $(0.8,1.0)$  &  $(0.3,0.4)$   &  $(0.0,0.0)$   &  $(0.0,0.0)$   \\
   \hline
   BCH $(15,5,3)$  & 0.333 &  $(4.1,4.6)$   &   $(2.7,2.9)$    &    $(2.0,2.1)$  &  $(1.5,1.60)$  &  $(0.7,0.8)$   &  $(0.2,0.2)$   \\
   \hline
   BCH $(31,26,1)$ & 0.839 &  $(1.2,1.5)$   &   $(0.2,0.2)$    &    $(0.0,0.0)$  &  $(0.0,0.0)$   &  $(0.0,0.0)$   &  $(0.0,0.0)$   \\
   \hline
   BCH $(31,21,2)$ & 0.677 &  $(2.3,2.9)$   &   $(1.7,2.0)$    &    $(0.7,0.8)$  &  $(0.2,0.2)$   &  $(0.0,0.0)$   &  $(0.0,0.0)$   \\
   \hline
   BCH $(31,16,3)$ & 0.516 &  $(2.9,3.1)$   &   $(2.1,2.2)$    &    $(1.5,1.6)$  &  $(1.3,1.4)$   &  $(0.6,0.7)$   &  $(0.1,0.1)$   \\
   \hline
   BCH $(31,11,5)$ & 0.355 &  $(4.1,4.4)$   &   $(3.5,4.2)$    &    $(2.2,2.3)$  &  $(2.0,2.1)$   &  $(1.8,2.0)$   &  $(1.1,1.3)$   \\
   \hline
   BCH $(31,6,7)$  & 0.194 &  $(5.4,5.9)$   &   $(4.3,4.8)$    &    $(3.5,3.8)$  &  $(3.2,3.3)$   &  $(2.7,2.8)$   &  $(2.3,2.4)$   \\
   \hline
   \hline
   Convolutional Code  & $R^{CC}_c$ & M=2  & M=4  & M=8  &  M=16  &  M=32  &  M=64\\
  \hline
  trel$(6,[53~~75])$    & 0.500 &   $(3.8,4.6)$  &   $(2.7,3.1)$  &    $(2.1,2.3)$  &  $(1.8,2.0)$   &  $(1.4,1.5)$   &  $(1.4,1.4)$   \\
  \hline
   trel$(7,[133~~171])$  & 0.500 &   $(4.0,4.7)$  &   $(3.0,3.5)$  &    $(2.2,2.4)$  &  $(1.8,2.0)$   &  $(1.5,1.6)$   &  $(1.4,1.5)$   \\
  \hline
  trel$(7,[133~~165~~171])$  & 0.333 &   $(5.7,6.4)$  &   $(4.8,5.1)$  &    $(3.7,3.9)$  &  $(3.1,3.3)$   &  $(2.7,2.8)$   &  $(2.5,2.6)$   \\
   \hline
   trel$([4~3],[4~5~17;7~4~2])$  & 0.667 &   $(2.2,2.6)$  &   $(1.5,1.7)$  &    $(0.9,1.1)$  &  $(0.6,0.6)$   &  $(0.5,0.5)$   &  $(0.5,0.5)$   \\
   \hline
   trel$([5~4],[23~35~0;0~5~13])$  & 0.667 &   $(2.9,3.5)$  &   $(1.9,2.4)$  &    $(1.4,1.8)$  &  $(1.1,1.2)$   &  $(0.8,0.9)$   &  $(0.7,0.8)$   \\
   \hline
   \end{tabular}
\end{table*}

\begin{figure}[t]
\centerline{\psfig{figure=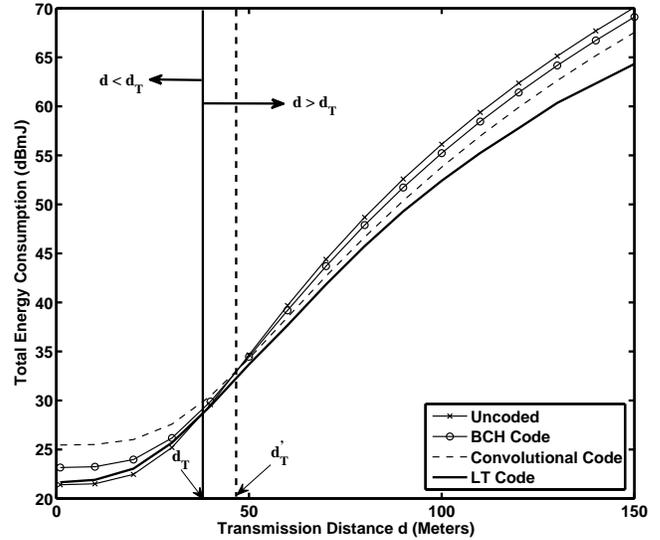,width=3.9in}}
\caption{Total energy consumption of optimized coded and uncoded
MFSK versus $d$ for $P_b=10^{-3}$.} \label{fig: Coded_Uncoded}
\end{figure}

Fig. \ref{fig: Coded_Uncoded} shows the total energy consumption
versus distance $d$ for the optimized BCH, convolutional and LT
coded MFSK schemes, compared to the optimized uncoded MFSK for
$P_b=10^{-3}$. The optimization is done over $M$ and the parameters
of coding scheme. Simulation results show that for $d$ less than the
threshold level $d_T \approx 40$ m, the total energy consumption of
optimized uncoded MFSK is less than that of the coded MFSK schemes.
However, the energy gap between LT coded and uncoded MFSK is
negligible compared to the other coded schemes as expected. For
$d>d_T$, the LT coded MFSK scheme is more energy-efficient than
uncoded and other coded MFSK schemes. Also, it is observed that the
energy gap between LT and convolutional coded MFSK increases when
the distance $d$ grows. This result comes from the high coding gain
capability of LT codes which confirms our analysis in Section
\ref{analysis_Ch4}. The threshold level $d_T$ (for LT code) or
$d^{'}_T$ (for BCH and convolutional codes) are obtained when the
total energy consumptions of coded and uncoded systems become equal.
For instance, using $\mathcal{L}_d =M_ld^\eta \mathcal{L}_1$, and
the equality between (\ref{energy_totFSK}) and
(\ref{energy_totalcodedFSK}) for uncoded and convolutional coded
MFSK, we have
\begin{equation}
d^{'}_T=\left[\hat{\Upsilon}_c^{CC}\dfrac{a_2
(1-\hat{R}_c^{CC})+N(\mathcal{E}^{CC}_{enc}+\mathcal{E}^{CC}_{dec})}{a_1(\hat{\Upsilon}_{c}^{CC}\hat{R}_c^{CC}-1)}\right]^{\frac{1}{\eta}},
\end{equation}
where
\begin{eqnarray}
\notag a_1 &\triangleq &(1+\alpha) \left[\left(
1-\left(1-\dfrac{2(M-1)}{M}P_b\right)^{\frac{1}{M-1}}\right)^{-1}-2
\right]\\
&\times&\dfrac{\mathcal{L}_1 M_l}{\log_2 M}\dfrac{N N_0}{\Omega},
\end{eqnarray}
\begin{equation}
a_2 \triangleq (\mathcal{P}_{c}-\mathcal{P}_{Amp})\dfrac{MN}{B
\log_2 M},
\end{equation}
\begin{equation}
(\hat{\Upsilon}_c^{CC},\hat{R}_c^{CC})=
\textrm{arg}~\min_{\Upsilon_c^{CC},R_c^{CC}}~\mathcal{E}^{CC}_{N,c}.
\end{equation}

 It should be noted that the above threshold level imposes a
constraint on the design of the physical layer of some wireless
sensor networking applications, in particular dynamic WSNs. To
obtain more insight into this issue, let assume that the location of
the sensor node is changed every $T_d \gg T_c$ time unit, where
$T_c$ is the channel coherence time. For the moment, let us assume
that the sensor node aims to choose either a \emph{fixed-rate coded}
or an uncoded MFSK based on the distance between sensor and sink
nodes. According to the results in Fig. \ref{fig: Coded_Uncoded}, it
is revealed that using fixed-rate channel coding is not energy
efficient for short distance transmission (i.e., $d < d^{'}_T$),
while for $d > d^{'}_T$, convolutional coded MFSK is more
energy-efficient than other schemes. For this configuration, the
sensor node must have the capability of an adaptive coding scheme
for each distance $d$. However, as discussed previously, the LT
codes can adjust their rates for each channel condition and have
(with a good approximation) minimum energy consumption for every
distance $d$. This indicates that LT codes can surpass the above
distance constraint for WSN applications with dynamic position
sensor nodes over Rayleigh fading channels. This characteristic of
LT codes results in reducing the complexity of the network design as
well. Of interest is the strong benefits of using LT coded MFSK
compared with the coded modulation schemes in
\cite{Cui_GoldsmithITWC0905, Chouhan_ITWC1009}. In contrast to
classical fixed-rate codes used in \cite{Cui_GoldsmithITWC0905,
Chouhan_ITWC1009}, the LT codes can vary their block lengths to
adapt to any channel condition in each distance $d$. Unlike
\cite{Cui_GoldsmithITWC0905} and \cite{Chouhan_ITWC1009}, where the
authors consider fixed-rate codes over an AWGN channel model, we
considered a Rayleigh fading channel which is a general model in
practical WSNs. The simplicity and flexibility advantages of LT
codes with an MFSK scheme make them the preferable choice for
wireless sensor networks, in particular for WSNs with dynamic
position sensor nodes.

\begin{figure}[t]
\centerline{\psfig{figure=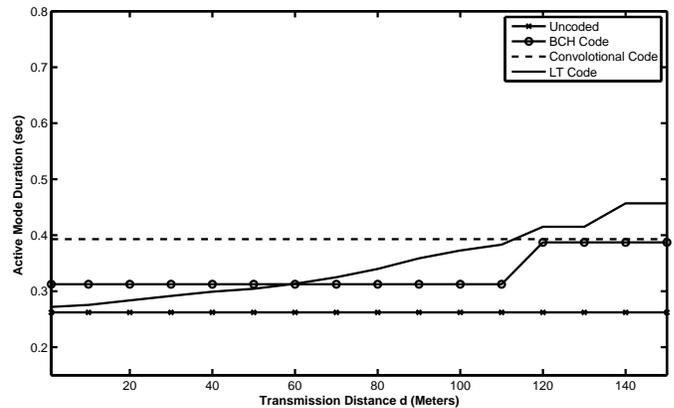,width=4.2in}}
\caption{Active mode duration of the optimized uncoded and coded MFSK schemes versus transmission distance $d$.} \label{fig: active}
\end{figure}

\textbf{Remark 1:} As discussed previously, the proposed LT coded scheme
benefits in adjusting the coding parameters in each channel realization
or equivalently each distance $d$ to minimize the total energy consumption.
As we will show shortly, this comes from a variable transmission time process
which adaptively controls the power consumption of the proposed scheme.
To address this issue, we plot the active mode duration of the optimized
LT coded modulation versus $d$ compared to those of the optimized uncoded
and other coded modulation schemes in Fig. \ref{fig: active}.
For this purpose, we use $T_{ac,c}=\frac{T_{ac}}{R_c}$, where $T_{ac}=\frac{MN}{B \log_2 M}$
denotes the active mode duration of uncoded scheme. Simulation results show
that the optimum constellation size $M$ that minimizes the total energy
consumption for the aforementioned schemes and for every value of distance
$d$ is $\hat{M}=2$. It is revealed from Fig. \ref{fig: active} that the
optimized uncoded and convolutional coded MFSK display fixed values of
$T_{ac}$ and $T_{ac,c}^{CC}$ for every value of $d$. In addition, we
can see that for $d < 60$ m and $d<110$ m, $T_{ac,c}^{LT}$ is less than
$T_{ac,c}^{BC}$ and $T_{ac,c}^{CC}$, respectively. Using (\ref{total_energy1})
and Fig. \ref{fig: Coded_Uncoded}, and noting that the total energy consumption of a certain scheme is proportional
to the transmission time, it is concluded that the scheme with a lower transmission time
is not necessarily more energy efficient than that of the scheme with greater $T_{ac}$.

\textbf{Remark 2:} To make a fair comparison between uncoded and coded schemes,
one would expect to assume a constant active mode duration for all the communication
schemes. According to $T_{ac,c}=\frac{T_{ac}}{R_c}$ with $T_{ac}=\frac{MN}{B \log_2 M}$,
this is achieved by adjusting the modulation order $M$. However, it is
worth mentioning that the assumption of the same transmission times for all the
schemes is not a realistic assumption in feasible WSNs where ``autonomous'' sensor
devices are powered by limited-lifetime batteries. More precisely, under the
assumption of the same $T_{ac}$ (or $T_{ac,c}$), the wireless sensor network
needs an extra hardware to adjust the constellation size $M$ (in each distance $d$),
which imposes more cost, complexity and power consumption in the network. While,
our scheme with variable transmission times surpasses the above adaptive
modulation constraint using the fact that the optimum constellation size $M$ which
minimizes the total energy consumption (in each distance $d$) is $\hat{M}=2$.

\section{Conclusion}\label{conclusion_Ch6}
In this paper, we analyzed the energy efficiency of LT coded MFSK in
a proactive WSN over Rayleigh fading channels with path-loss. It was
shown that the energy efficiency of LT codes is similar to that of
uncoded MFSK scheme for $d<d_T$, while for $d > d_T$, LT coded MFSK
outperforms other uncoded and coded schemes, from the energy
efficiency point of view. This result follows from the flexibility
of the LT code to adjust its rate and the corresponding LT coding
gain to suit instantaneous channel conditions for any transmission
distance $d$. This rate flexibility offers strong benefits in using
LT codes in practical WSNs with dynamic distance and position
sensors. In such systems and for every value of distance $d$, LT
codes can adjust their rates to achieve a certain BER with low
energy consumption. The importance of our scheme is that it avoids
some of the problems inherent in adaptive coding or Incremental
Redundancy (IR) systems (channel feedback, large buffers, or
multiple decodings), as well as the coding design challenge for
fixed-rate codes used in WSNs with dynamic position sensor nodes.
The simplicity and flexibility advantages of LT codes make the LT
code with MFSK modulation can be considered as a \emph{Green
Modulation/Coding} (GMC) scheme in dynamic WSNs.

In this paper, we have shown the significant benefit of rateless codes
(with focus on the optimized LT codes) for sensor networks over the more
traditional fixed-rate codes. Our future research involves selecting the
best rate-adaptive code among LT, Raptor, punctured LDPC, and punctured
Turbo codes; in particular, we are interested to study the performance
of the Raptor code which has a linear-time encoding versus the non-linear
cost of the LT code. For this study, a particularly nice feature of the
LT code is to rapidly optimize the code using a modified EXIT chart
strategy introduced in \cite{Etesami_ITIT0506}.

\appendices

\section{Asymptotic LT Code Rate}
The proof of the remark is straightforward using the notation of
\cite{Etesami_ITIT0506} and the bipartite graph concepts in graph
theory. Obviously, the output-node degree distribution
$\mathcal{O}(x)$ induces a distribution on the input nodes in the
bipartite graph. Thus, in the asymptotic case of $k \to \infty$, we
have the input-node degree distribution defined as $\mathcal{I}(x)
\triangleq \sum_{i=1}^{\infty}\nu_i x^i$, where $\nu_i$ denotes the
probability that an input node has a degree $i$. In this case, the
average degree of the input and output nodes are computed as
$\frac{d \mathcal{I}(x)}{d x}\big|_{x=1} \triangleq
\mathcal{I}_{ave}$ and $\frac{d \mathcal{O}(x)}{d x}\big|_{x=1}
\triangleq \mathcal{O}_{ave}$, respectively. Thus, the number of
edges exiting the input nodes of the bipartite graph, in the
asymptotic case of $k \to \infty$, is $k\mathcal{I}_{ave}$, which
must be equal to $n\mathcal{O}_{ave}$, the number of edges entering
the output nodes in the graph. As a results, the asymptotic LT code
rate is obtained as $R^{LT}_c= \frac{k}{n} \approx
\frac{\mathcal{O}_{ave}}{\mathcal{I}_{ave}}$, which is a
deterministic value for given $\mathcal{O}(x)$ and $\mathcal{I}(x)$.


\begin{biography}[{\includegraphics[width=1in,height=1.25in,clip,keepaspectratio]{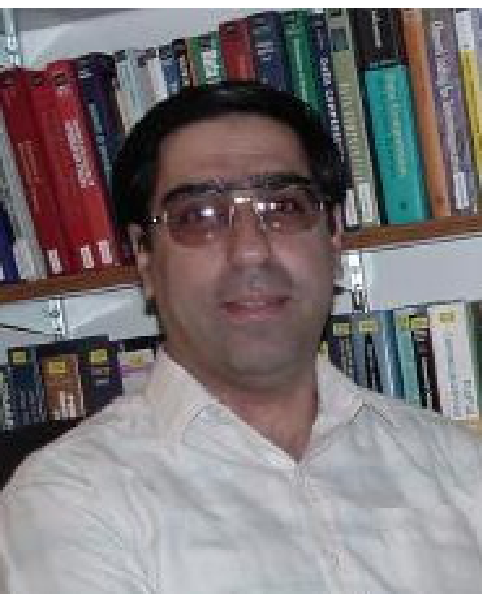}}]
{Jamshid Abouei} received the B.Sc. degree in electronics engineering and
the M.Sc. degree in communication systems engineering (with the highest
honor) both from the Isfahan University of Technology (IUT), Iran, in 1993
and 1996, respectively, and the Ph.D. degree in electrical engineering from
the University of Waterloo in Waterloo, ON, Canada, in 2009.

From 1996 to 2004, he was a faculty member (lecturer) in the Department of
Electrical Engineering, Yazd University, and from 1998 to 2004, he was a
technical advisor and design engineer (part-time) in R\&D center and cable
design department in SGCC company. From 2009 to 2010, he was a Postdoctoral
Fellow in the Multimedia Lab, in the Department of Electrical \& Computer
Engineering, at the University of Toronto, ON, Canada.

Currently, Dr Abouei is an Assistant Professor in the Department of Electrical
\& Computer Engineering, at the Yazd University, Iran. His research interests
are in general areas of wireless ad hoc and sensor networks, with particular
reference to energy efficiency and optimal resource allocation, multi-user
information theory, cooperative communication in wireless relay networks,
applications of game theory, and orthogonal
codes in CDMA systems.

Dr Abouei has received numerous awards and scholarships, including FOE and IGSA
awards for excellence in research in University of Waterloo, Canada, and MSRT Ph.D.
Scholarship from the Ministry of Science, Research and Technology, Iran in 2004.
\end{biography}

\begin{biography}[{\includegraphics[width=1in,height=1.25in,clip,keepaspectratio]{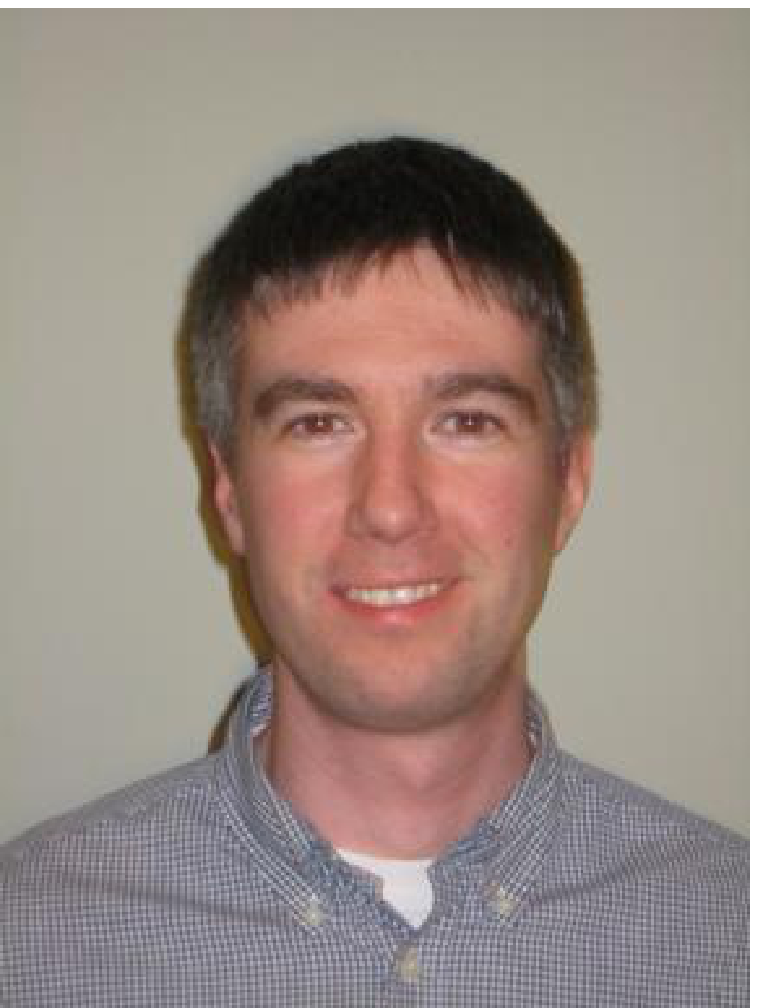}}]
{J. David Brown} was born in Ottawa, ON, Canada, in 1977.  He received the B.Sc.(Eng.)
degree in electrical and computer engineering in 2000, and the M.Sc.(Eng.) degree in
2002, both from Queen's University in Kingston, ON, Canada.  In 2008, he received the
Ph.D. degree in electrical and computer engineering from the University of Toronto in
Toronto, ON, Canada.

From 2002 to 2004 and again from 2008 to 2009, he worked as an Electrical Engineer at
General Motors.  In 2009, he joined the Network Information Operations Section at DRDC
in Ottawa, ON, Canada, as a Research Scientist.  His research interests include digital
communications, error-control codes, and machine learning.

Dr. Brown has received numerous awards and scholarships, including the Natural Sciences
and Engineering Research Council of Canada (NSERC) Post-graduate Scholarship, the NSERC
Canada Graduate Scholarship (CGS), and two Industry Canada Fessenden Postgraduate Scholarships.
He also received the Queen's University Professional Engineers of Ontario Gold Medal.
\end{biography}

\begin{biography}[{\includegraphics[width=1in,height=1.25in,clip,keepaspectratio]{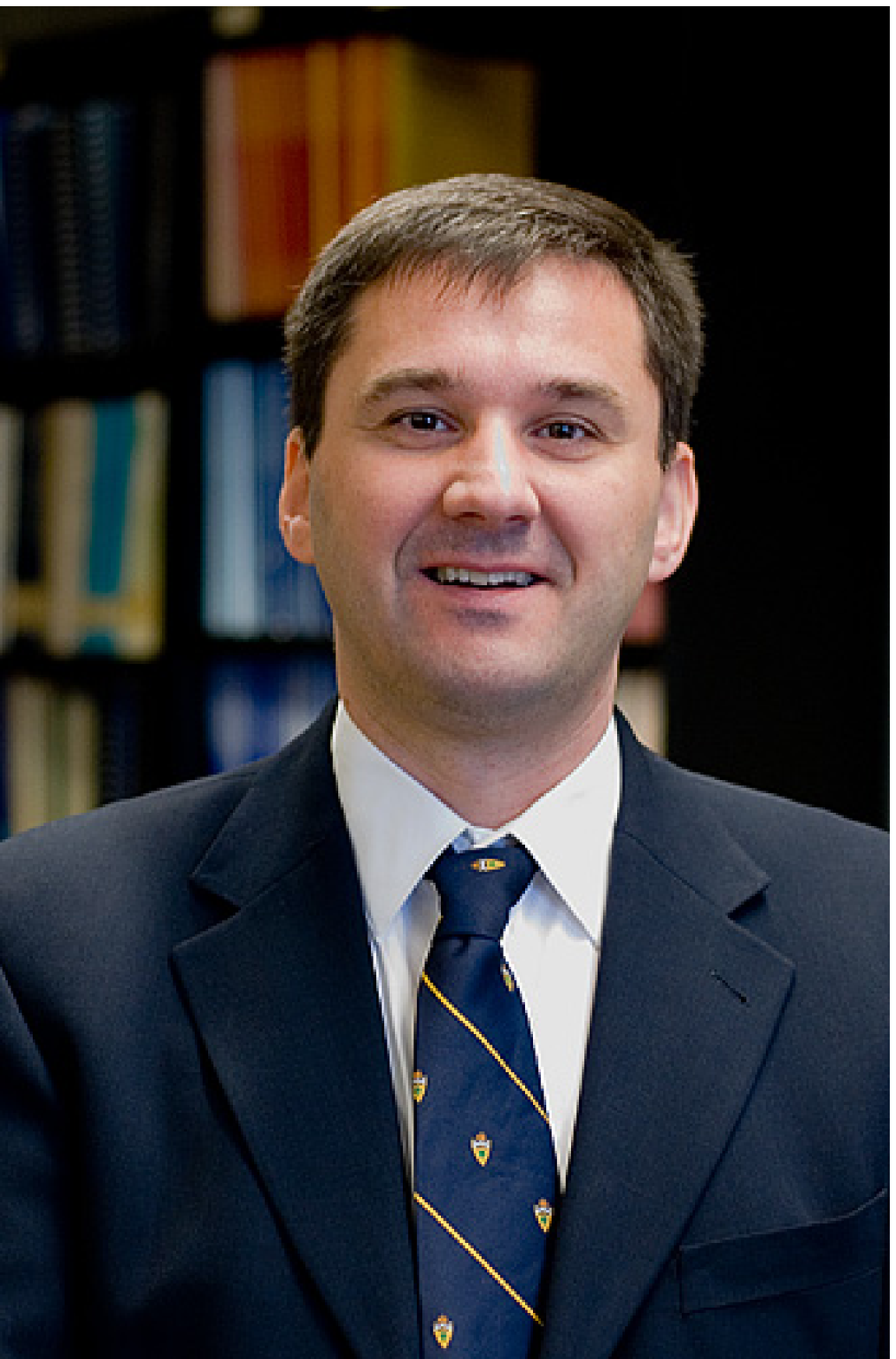}}]
{Konstantinos N. (Kostas) Plataniotis} is a Professor with the
Edward S. Rogers Sr. Department of Electrical and Computer
Engineering at the University of Toronto in Toronto, Ontario,
Canada, and an Adjunct Professor with the School of Computer
Science at Ryerson University, Canada. He is the Director of The
University of Toronto's Knowledge Media Design Institute
(www.kmdi.utoronto.ca), and the Director of Research for the
Identity, Privacy and Security Institute at the University of
Toronto (www.ipsi.utoronto.ca).

Prof. Plataniotis is the Editor in Chief (2009-2011) for the IEEE
Signal Processing Letters and chairs the Examination Committee for
the IEEE Certified Biometrics Professional (CBP) Program
(www.ieeebiometricscertification.org). He served on the IEEE
Educational Activities Board (EAB) and he
was the Chair (2008-09) of the IEEE EAB Continuing Professional
Education Committee. Dr. Plataniotis has served as Chair
(2000-2002) IEEE Toronto Signal Processing Chapter, Chair
(2004-2005) IEEE Toronto Section, and he was a member of the 2006
and 2007 IEEE Admissions \& Advancement Committees.

He is the 2005 recipient of IEEE Canada's Outstanding Engineering Educator Award
``for contributions to engineering education and inspirational guidance of graduate students'' and the co-recipient of the
2006 IEEE Trans. on Neural Networks Outstanding Paper Award for the published in 2003 paper
entitled `` Face Recognition Using Kernel Direct Discriminant Analysis Algorithms''.

He is a registered professional engineer in the province of
Ontario, and a member of the Technical Chamber of Greece, and a Fellow of the Engineering Institute of Canada.

His research interests include biometrics, communications systems,
multimedia systems, and signal \& image processing.

\end{biography}

\begin{biography}[{\includegraphics[width=1in,height=1.25in,clip,keepaspectratio]{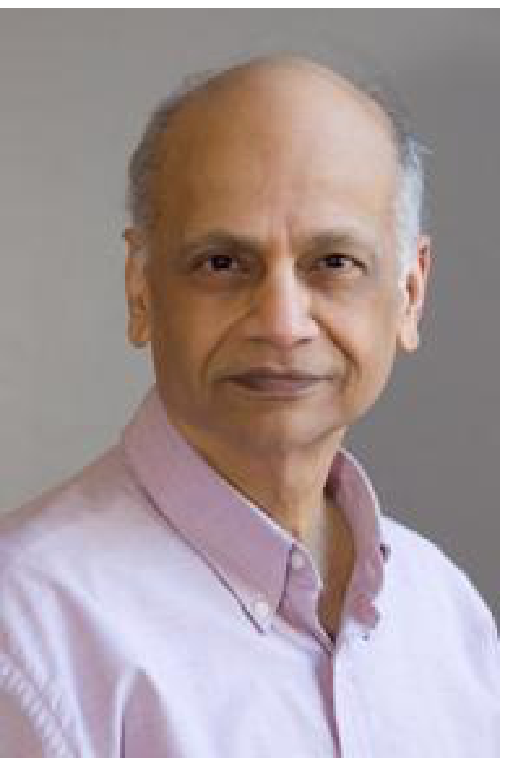}}]
{Subbarayan Pasupathy} was born in Chennai (Madras), Tamilnadu,
India. He received the B.E. degree in telecommunications from the
University of Madras, the M.Tech. degree in electrical engineering
from the Indian Institute of Technology, Madras, and the M.Phil. and
Ph.D. degree in engineering and applied science from Yale University.

Currently, he is a Professor Emeritus in the Department of Electrical
and Computer Engineering at the University of Toronto, where he has
been a Faculty member from 1972. His research over the last three
decades has mainly been in statistical communication theory and
signal processing and their applications to digital communications.
He has served as the Chairman of the Communications Group and as the
Associate Chairman of the Department of Electrical Engineering at the
University of Toronto. He is a registered Professional Engineer in the
province of Ontario. During 1982-1989 he was an Editor for {\it Data
Communications and Modulation} for the IEEE TRANSACTIONS ON COMMUNICATIONS.
He has also served as a Technical Associate Editor for the IEEE COMMUNICATIONS
MAGAZINE (1979-1982) and as an Associate Editor for the {\it Canadian
Electrical Engineering Journal} (1980-1983). He wrote a regular humour column
entitled ``Light Traffic'' for the IEEE COMMUNICATIONS MAGAZINE during 1984-98.

Dr. S. Pasupathy was elected as a Fellow of the IEEE in 1991 ``for contributions
to bandwidth efficient coding and modulation schemes in digital communication'',
was awarded the Canadian Award in Telecommunications in 2003 by the {\it
Canadian Society of Information Theory}, was elected as a Fellow
of the {\it Engineering Institute of Canada} in 2004 and as a Fellow of the
{\it Canadian Academy of Engineering} in 2007. He was honoured as a
Distinguished Alumnus by I.I.T, Madras, India in 2010. He has been
identified as a ``highly cited researcher'' by ISI Web of Knowledge and
his name is listed in ISIHighlyCited.com.
\end{biography}

\end{document}